\documentclass[5p, sort&compress]{elsarticle} 
\usepackage{graphicx}
\usepackage{amssymb}
\usepackage{epstopdf}
\usepackage[amssymb,thinqspace,thinspace]{SIunits}
\usepackage{amsmath}
\usepackage{booktabs}
\usepackage{rotating}
\usepackage{multirow}
\usepackage{psfrag}
\usepackage{lineno}
\usepackage{textgreek}
\usepackage{lineno}
\modulolinenumbers[5]
\usepackage{color}
\usepackage{dblfloatfix} 

\journal{arXiv}

\begin{document}
\begin{frontmatter}
\title{Strengthening $\kappa$-carbide steels using residual dislocation content}
\author[1]{T. W. J. Kwok}
\author[1]{K. M. Rahman}
\author[1,2]{V. A. Vorontsov}
\author[1]{D. Dye\corref{cor1}}\ead{ddye@ic.ac.uk}
\cortext[cor1]{Corresponding author}
\address[1]{Department of Materials, Royal School of Mines, Imperial College London, Prince Consort Road, London SW7 2BP, UK}
\address[2]{Design, Manufacturing and Engineering Management, James Weir Building, University of Strathclyde, 75 Montrose St, Glasgow G1 1XJ, UK}

\begin{abstract}


A steel with nominal composition Fe-28Mn-8Al-1.0C in mass percent was hot rolled at two temperatures, 1100 \degree C and 850 \degree C and subsequently aged at 550 \degree C for 24 h. The lower temperature rolling resulted in a yield strength increment of 299 MPa while still retaining an elongation to failure of over 30\%. The large improvement in strength was attributed to an increase in residual dislocation density which was retained even after the ageing heat treatment. A homogeneous precipitation of $\kappa$-carbides in both samples also showed that the high residual dislocation density did not adversely affect precipitation kinetics. These findings demonstrate that the tensile properties of this class of steel can yet be improved by optimising hot rolling process parameters.

\end{abstract}

\end{frontmatter}



The development of advanced high strength steels with superior mechanical properties combined with a low density is a highly sought after asset for commercial steelmakers. Reducing the weight of structural components particularly in the automotive sector leads to better performance, reduced fuel consumption and lower emissions \cite{Rana2016}. This can be achieved by either reducing the gauge thickness of conventional engineering steels or by reductions in density \cite{Bouaziz2013}.

A new generation of low density alloys based on the Fe-Mn-Al-C system have received extensive research interest in recent years \cite{Chen2017,Raabe2014}. This class of steels exhibit yield strengths of between 400-1000$\usk\mega\pascal$ combined with an elongation to failure of above 60\% \cite{Gutierrez-Urrutia2014,Chen2017,DeCooman2018}. Within the Fe-Mn-Al-C system, a subgroup of steels with a typical composition range of Fe-(25-30)Mn-(6-10)Al-(0.8-2.0)C, are able to precipitate a large volume fraction of coherent $\kappa$-carbides in an austenitic matrix in a manner that resembles $\gamma/\gamma'$ microstructures in Ni-base superalloys \cite{Frommeyer2006,Chen2020}. These steels are typically hot rolled and quenched from temperatures $\geqslant$ 1100 \degree C \cite{Lin2014,Gutierrez-Urrutia2013a,Gutierrez-Urrutia2014,Yao2017} in order to facilitate the spinodal decomposition of high temperature austenite into solute (Al and/or C) lean and rich $\gamma'$ and $\gamma''$ austenite phases respectively, where the $\gamma''$ phase eventually forms the non-stoichiometric L$'$1\textsubscript{2} (Fe,Mn)\textsubscript{3}AlC\textsubscript{$x$} $\kappa$-carbide during ageing \cite{Zhang2021,Cheng2015a}. A high rolling temperature maintains a fully austenitic microstructure, preventing the formation of ferrite which negatively affects ductility \cite{Yoo2009}. After rolling, it is important that the steel must be quenched at a sufficiently high rate to prevent the uncontrolled precipitation of $\kappa$-carbides at undesirable locations such as grain boundaries where they cause embrittlement \cite{Kim2013,Liu2018}.

While these $\kappa$-carbide strenghtened steels have already been shown to possess high strengths and elongations, this study aims to investigate if they can be strengthened further while retaining an appreciable amount of ductility by process optimisation.


The nominal chemical composition of the steel was Fe-28Mn-8Al-1C in mass percent. The alloy was melted in a vacuum arc melter under an Ar atmosphere and cast into a 23$\times$23$\times$60 mm ingot. The ingot was then sectioned into bars with dimensions of 10$\times$10$\times$60$\usk\milli\meter$. The bars were homogenised at 1200 \degree C for 24 h in a quartz encapsulated tube backfilled with Ar and water quenched. Two bars were then hot rolled at either 1100$\celsius$ and 850$\celsius$ each to a total reduction of approximately 80\%. The rolled strips were returned to the furnace for 5 min after the final pass then water quenched. The strips were then aged at 550 \degree C for 24 h. Tensile samples with gauge dimensions of 19$\times$1.5$\times$1.5$\usk\milli\meter$ were obtained from each rolled strip \textit{via} electric discharge machining such that the tensile direction was parallel to the rolling direction. Tensile testing was performed using a Zwick 100$\usk\kilo\newton$ load frame at a nominal strain rate of $10^{-3}$ s\textsuperscript{-1}. Electron backscatter diffraction (EBSD) and Secondary Electron Microscopy (SEM) were performed on a Zeiss Auriga FEG-SEM equipped with a Bruker EBSD detector. Samples for Transmission Electron Microscopy (TEM) were obtained by lifting out a foil using a FEI Helios Nanolab 600 Focused Ion Beam (FIB) workstation. TEM analysis to obtain high resolution lattice images was conducted on an FEI Titan 80/300 TEM/STEM microscope, fitted with a monochromator and image aberration corrector.


\begin{figure}[t]
	\centering
	\includegraphics[width=\linewidth]{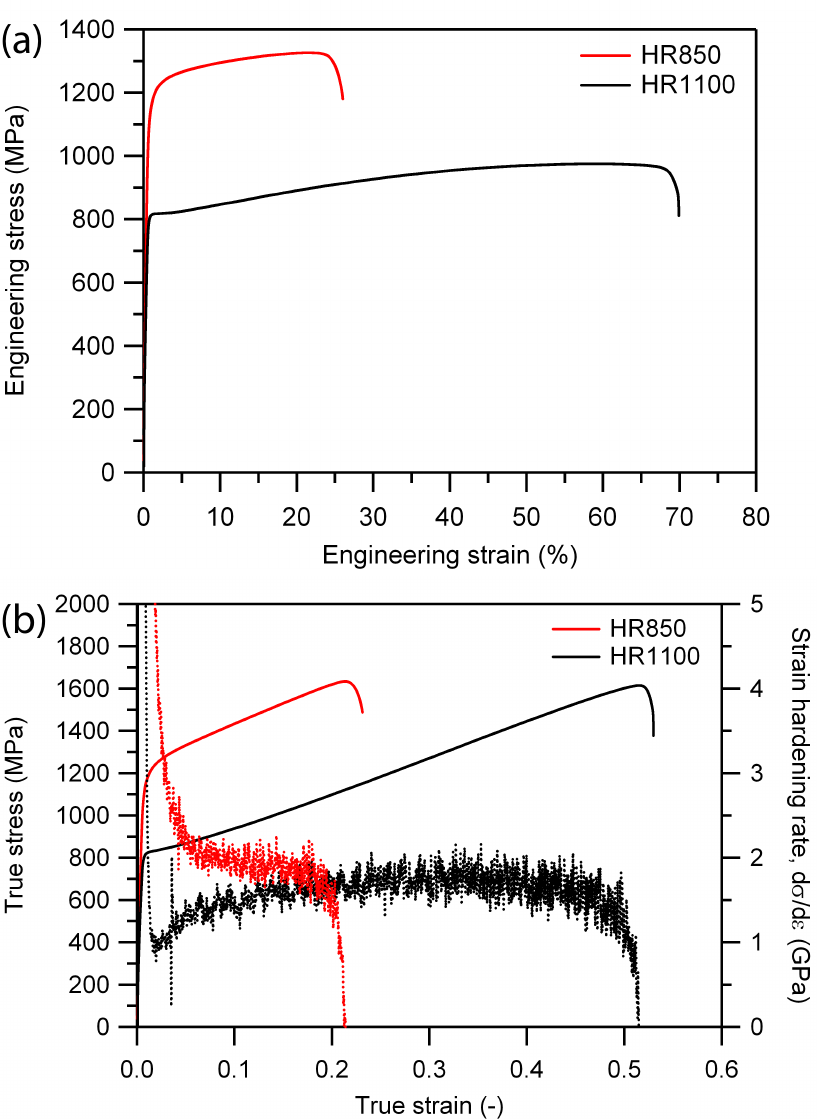}
	\caption{(a) Engineering tensile curves, (b) true stress-strain curves and the hardening rates (dashed lines) of HR1100 and HR850 samples.}
	\label{fig:kappa-tensiles}
\end{figure}

\begin{figure}[t!]
	\centering
	\includegraphics[width=\linewidth]{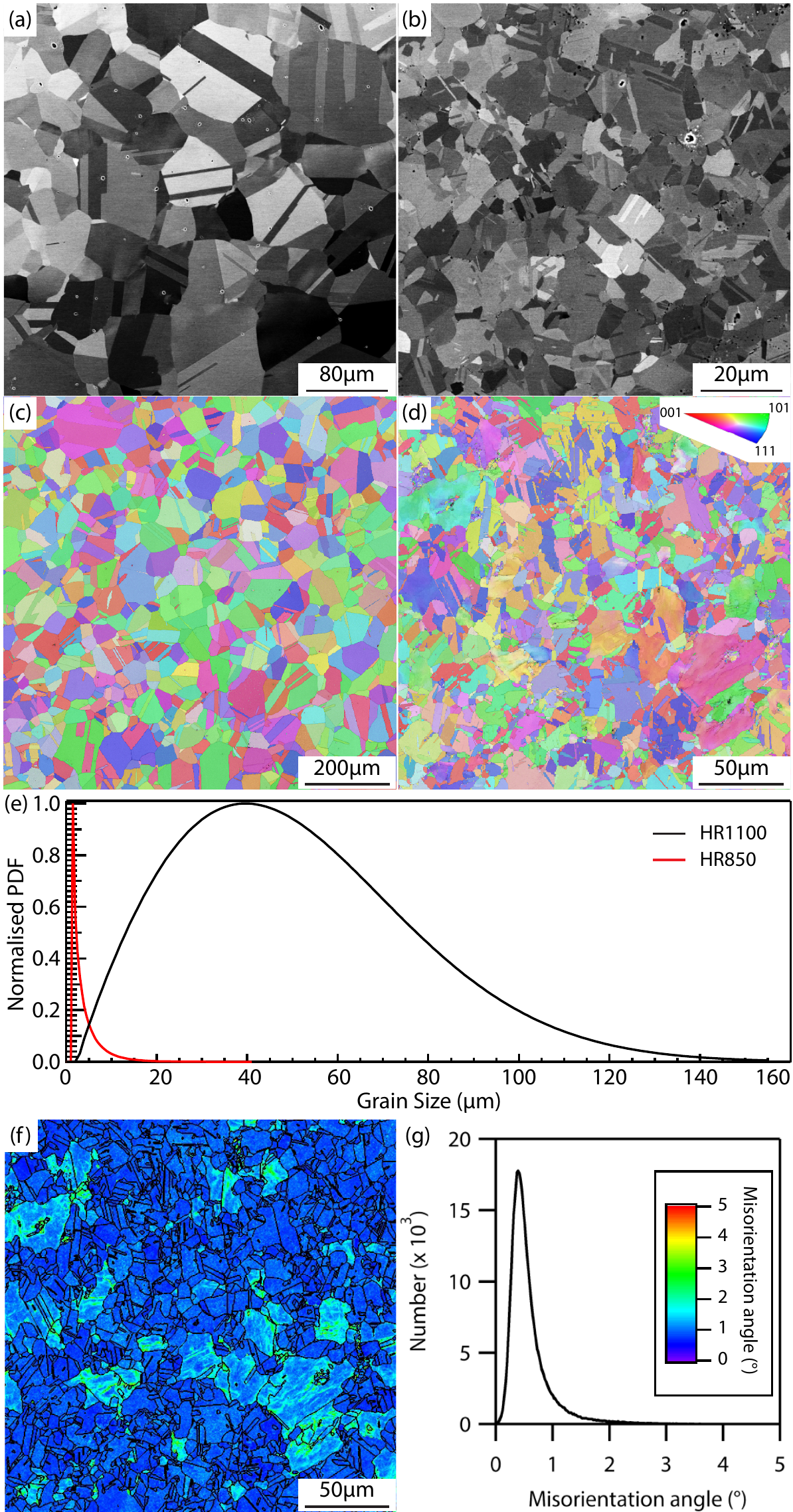}
	\caption{Backscatter electron micrograph of (a) HR1100 and (b) HR850 samples. EBSD IPF-X (right) maps of (c) HR1100 and (d) HR850. Strip normal direction is perpendicular to the page. (e) Normalised grain size Probability Distribution Function (PDF) obtained from (c) and (d). (f) KAM map of HR850, black lines indicate high angle and $\Sigma$3 boundaries, (g) misorientation angle frequency and colour legend of (f). }
	\label{fig:armicro}
\end{figure}

\begin{table}
	\centering
	\caption{Summary of tensile properties of the investigated steels.}
	\begin{tabular}{lcccc}
		\toprule
		& \textit{E}     & $\sigma_{0.2}$ & UTS   & $\varepsilon_f$\\
		& (GPa) & (MPa) & (MPa) & (\%) \\
		\midrule
		HR850 & 1.70  & 1115  & 1326  & 26 \\
		HR1100 & 1.10  & 816   & 975   & 70 \\
		\bottomrule
	\end{tabular}%
	\label{tab:mech props}%
\end{table}%

Tensile results are shown in Figure \ref{fig:kappa-tensiles} where HR1100 and HR850 refer to the strips rolled at 1100 \degree C and 850 \degree C respectively. A summary of the tensile properties is shown in Table \ref{tab:mech props}. The HR1100 sample possessed a lower yield strength but a large elongation to failure, whereas the HR850 sample had a significantly higher yield strength ($+299$ MPa) but with a much lower elongation ($-42$\%) compared to HR1100. Nevertheless, it should be noted that HR850 still possessed a total elongation of 26\% which is comparable to many automotive steels \cite{Rana2016,Keeler2014}. 

Another observation was that both steels failed at approximately the same true UTS, suggesting that both steels have reached a similar Strain Hardening Rate (SHR) and subsequently failed by necking \textit{i.e.} $\sigma = d\sigma/d\varepsilon$. This also gives confidence that the small geometry of the tensile specimens used did not result in premature failure.

The HR850 sample also showed a much higher initial SHR but decreased steadily up to failure whereas the HR1100 sample showed a lower initial SHR which gradually increased to 1.6 GPa, remained fairly constant and then began to decrease before eventual failure.

\begin{figure*}[t!]
	\centering
	\includegraphics[width=\linewidth]{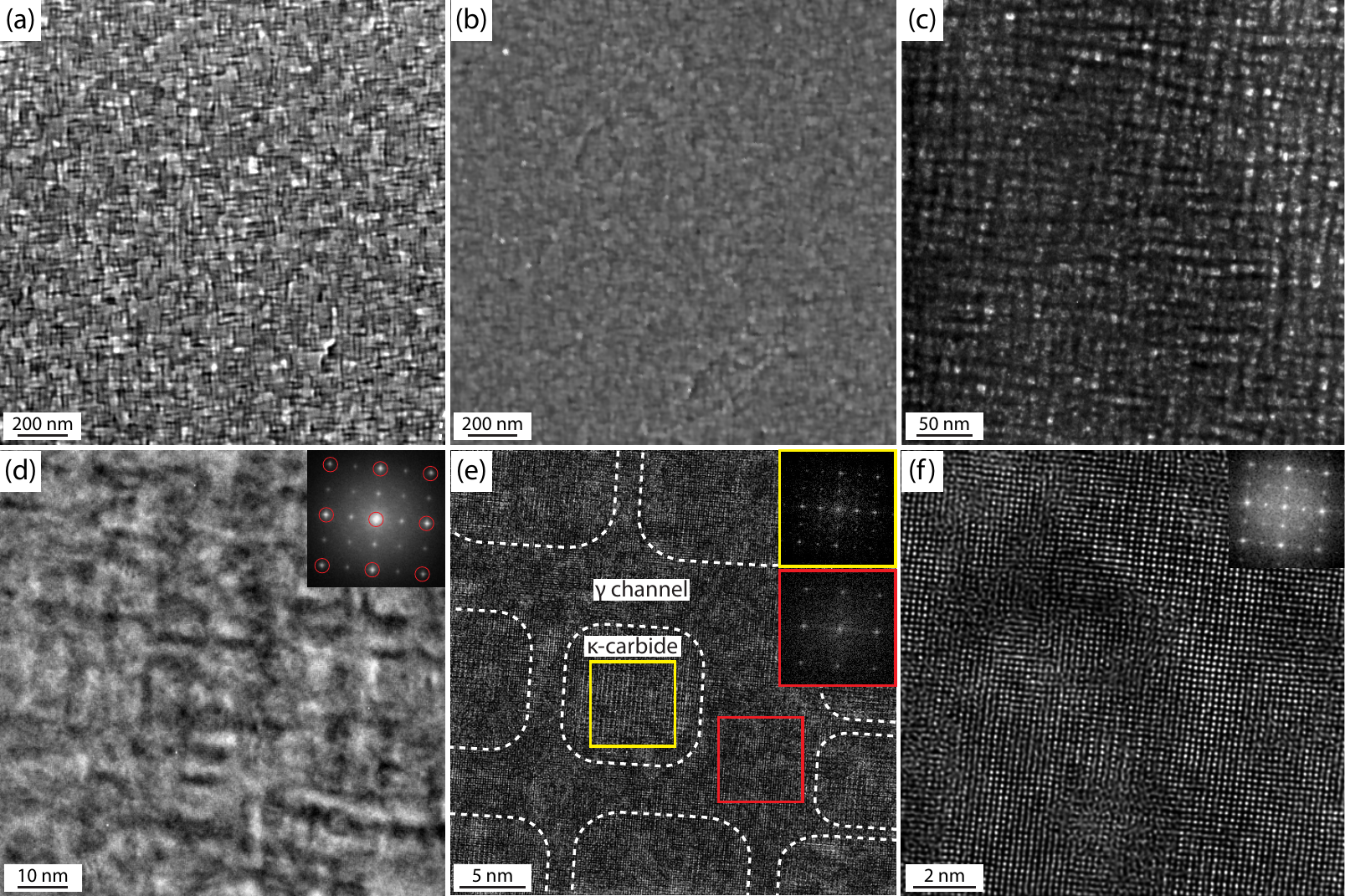}
	\caption{Etched secondary electron micrographs of (a) HR1100 and (b) HR850. (c) TEM-DF micrograph of $\kappa$-carbides in HR1100. (d-f) HR-TEM micrographs of $\kappa$-carbides and $\gamma$ matrix in the HR1100 sample at increasing magnifications. Insets: Fast Fourier Transform (FFT) of respective micrographs. (d) Inset: $[001]_\gamma$ pseudo-diffraction pattern circled in red. (e) FFT from the yellow and red squares showing $\kappa$-carbide and $\gamma$ pseudo-diffraction patterns respectively. Positions of $\kappa$-carbides are drawn based on local FFT. (f) HR-TEM micrograph of the whole $\kappa$-carbide. Electron beam direction parallel to $\langle 001 \rangle$ for (c-f).}
	\label{fig:kappa-ppts}
\end{figure*}

The microstructures of HR1100 and HR850 are shown in Figure \ref{fig:armicro}. Both samples were found to be fully austenitic; ferrite and coarse intergranular $\kappa$-carbides were not observed. The HR1100 sample possessed a significantly larger grain size compared to HR850. Based on the EBSD data in Figures \ref{fig:armicro}c-d, it can be seen that the HR1100 sample also possessed a wide distribution of grain sizes with an area weighted mean of 72.2 \textmu m while the HR850 sample had a much narrower grain size distribution with an area weighted mean of 14.2 \textmu m. This shows that a lower rolling temperature was able to significantly reduce the grain size as well as generate a more uniform grain size distribution. It is worth noting that a significant residual dislocation density was present in the HR850 sample as shown in the Kernel Average Misorientation (KAM) map in Figures \ref{fig:armicro}f-g.

In order to investigate if rolling at a lower temperature and the residual dislocation density in the HR850 sample may have affected the $\kappa$-carbide precipitation behaviour, both HR1100 and HR850 samples were polished and etched in nital (2\% HNO\textsubscript{3} in ethanol) and imaged in an SEM. Figures \ref{fig:kappa-ppts}a-b show a homogeneous distribution of intragranular $\kappa$-carbides in both HR1100 and HR850 samples respectively. The HR850 sample did not show evidence of $\kappa$-carbide rafting, which is known to deletriously affect tensile properties in steels aged above 800 \degree C \cite{Lin2010,Cheng2015a}.

The $\kappa$-carbides in the HR1100 sample were imaged in a TEM (Figures \ref{fig:kappa-ppts}c-f), revealing an average carbide size of $7.5 \pm 0.9$ nm and an approximate volume fraction of 0.3. The austenite channels between each $\kappa$-carbide are also very thin, typically finer or as fine as the $\kappa$-carbide diameter. It is acknowledged that the measurement of $\kappa$-carbide size and volume fraction from the 2D projection of the $\langle 100 \rangle$ zone may be prone to error due to the 3D network structure of $\kappa$-carbide cubes and plates \cite{Yao2017,Yao2016,Bartlett2014}. The $\kappa$-carbide size in the HR850 sample was estimated to be approximately 9 nm based on SEM micrographs at higher magnifications. The small difference in $\kappa$-carbide sizes between HR1100 and HR850 suggests that any apparent size differences in Figure \ref{fig:kappa-ppts}a-b may be due to an etching effect with nital.

In order to determine the reason behind the higher yield strength in the HR850 sample, the classical strengthening equation can be invoked where the total yield strength, $\sigma_y$, is the sum of its individual contributing elements:

\begin{equation}
\label{eq:strength}
\sigma_y = \sigma_0 + \sigma_{\text{HP}} + \sigma_{ppt}+\sigma_{disl}
\end{equation} 

\noindent
where $\sigma_0$ is the sum of the inherent lattice strength and solid solution strength and should be equal between both HR1100 and HR850. $\sigma_{\text{HP}}$ is the Hall-Petch grain size strength and is equal to $k/\sqrt{d}$, where $k$ is the Hall-Petch parameter (taken to be 357 MPa \textmu m\textsuperscript{-0.5} \cite{Sevsek2018}) and $d$ is the grain diameter. $\sigma_{ppt}$ is the strength due to the $\kappa$-carbides and $\sigma_{disl}$ is the strength due to any additional plastic strain. 

The calculation of $\sigma_{ppt}$ is complex due to the 3D network structure and ordered nature of the $\kappa$-carbides. However, an equation with good accuracy for $\sigma_{ppt}$ due to $\kappa$-carbides as given by Yao \textit{et al.} \cite{Yao2017} can be shown as:

\begin{equation}
\label{eq:ppt_strength}
\small
\Delta \sigma_{ppt} = \frac{M}{N} \sqrt{\frac{3}{2}} \left(\frac{Gb}{r}\right) \sqrt{V_f} \frac{2w}{\sqrt{\pi^3}} \left(\frac{2\pi\gamma_{\text{\tiny APB}}r}{w G b^2}-1\right)^{0.5}
\end{equation}

\noindent
where $M$ is the Taylor factor taken to be 3.06, $G$ is the shear modulus taken to be $70\times10^3$ MPa, $b$ is the Burgers vector length taken to be 0.26 nm, $w$ is a dimensionless constant equal to 1, $r$ is the average $\kappa$-carbide radius, $V_f$ is the volume fraction of $\kappa$-carbides. $N$ is the number of dislocations in a pile-up ahead of a $\kappa$-carbide ranging from 4-8, here taken to be 6. $\gamma_{\text{\tiny APB}}$ is the antiphase boundary energy associated with shearing the ordered $\kappa$-carbide ranging from 350-700 mJ m\textsuperscript{-2}, here taken to be 350 mJ m\textsuperscript{-2} \cite{Yao2017}.

Using the average grain sizes obtained in Figure \ref{fig:armicro} as well as the average $\kappa$-carbide sizes and volume fractions obtained in Figure \ref{fig:kappa-ppts}, the individual elements in Equation \ref{eq:strength} can be determined and are shown in Table \ref{tab:strength_equation}. It can be seen that although the average grain size was reduced from 72.2 \textmu m to 14.2 \textmu m, the grain size strengthening increment was only a modest 53 MPa. This suggests that grain size reduction is not a very effective method of strengthening these steels unless the grain size can be reduced to below 5 \textmu m ($\sigma_{\text{HP}}=160$ MPa), which may not be feasible on an industrial scale. Recent efforts have been made to develop a similar $\kappa$-carbide strengthened Mn-lean duplex ($\gamma+\alpha$) steel in order to refine and stabilise a fine grain size \cite{Sohn2014}. However, it has been shown that precipitating a homogeneous distribution of $\kappa$-carbides in these alloys is difficult \cite{Sohn2013,Moon2018, Lu2015}.

\begin{table}[t]
	\centering
	\caption{Different contributors to yield strength in HR1100 and HR850 samples. }
	\begin{tabular}{lccccc}
		\toprule
		& $\sigma_0$ & $\sigma_{\text{HP}}$ & $\sigma_{ppt}$ & $\sigma_{disl}$ & $\sigma_{0.2}$ \\
		\midrule
		HR1100 & 274   & 42   &  514   & $-$ & 816 \\
		HR850 & 274   & 95    &   519  & 241 & 1115 \\
		\bottomrule
	\end{tabular}%
	\label{tab:strength_equation}%
\end{table}%

 Precipitation strengthening was nearly equal in both HR1100 and HR850, suggesting that small changes in $\kappa$-carbide size at a similar volume fraction did not lead to significant changes in strength. The largest contributor to strength, after subtracting away $\sigma_{\text{HP}}$ and $\sigma_{ppt}$, is therefore $\sigma_{disl}$ which was worth 241 MPa in HR850. This is unsurprising as strain gradients could be observed in the HR850 IPF X map as well as in the KAM map (Figure \ref{fig:armicro}d, f-g). It was surprising however, that the presence of a significant dislocation density necessary to raise the strength by 241 MPa was still retained even after the long ageing heat treatment. The dislocation density, $\rho$, contributing to $\sigma_{disl}$ may be estimated by using the following equation \cite{Lee2014}:

\begin{equation}
\sigma_{disl} = A M Gb\sqrt{\rho}
\end{equation}

\begin{figure}[t]
	\centering
	\includegraphics[width=\linewidth]{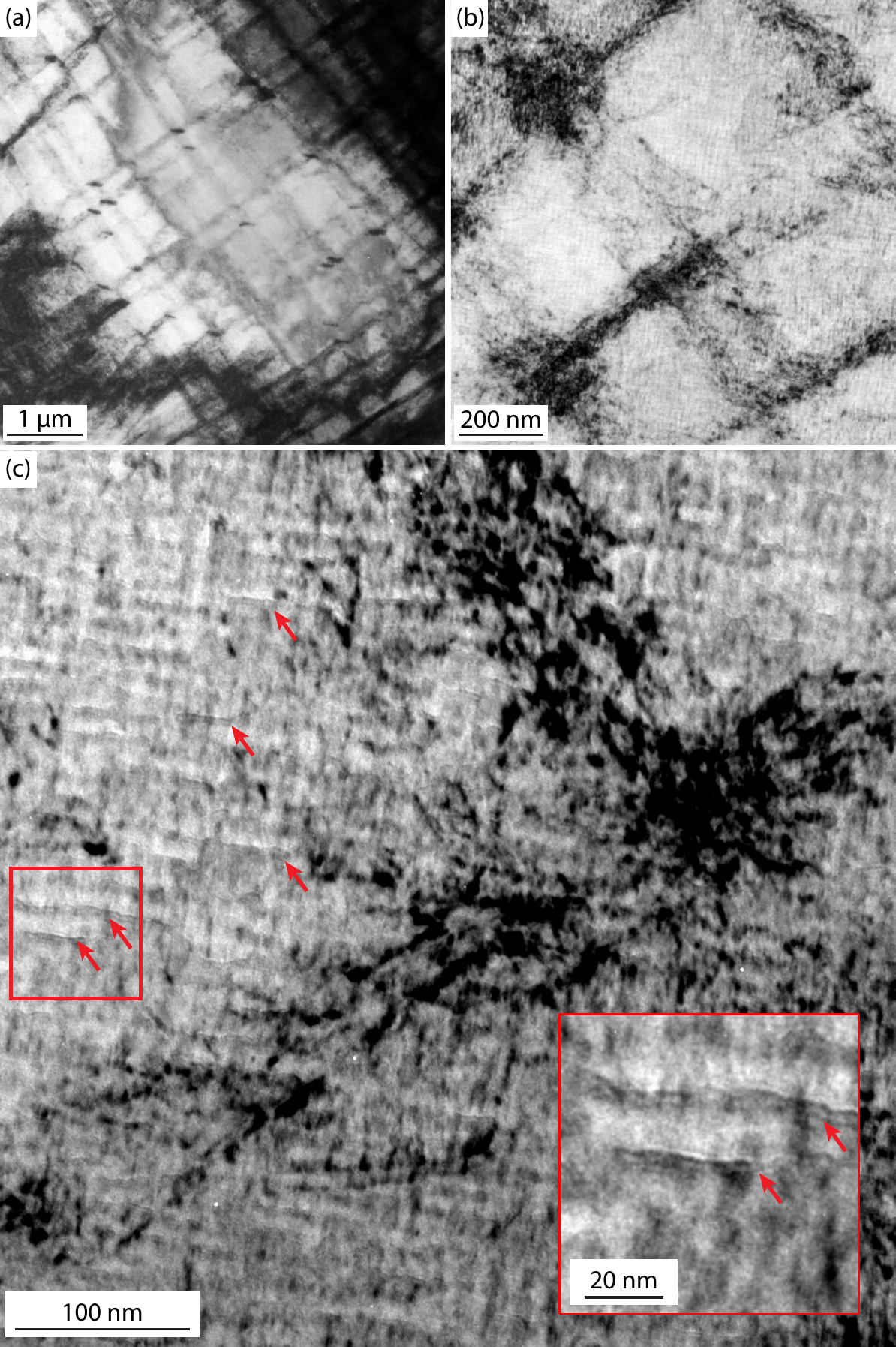}
	\caption{BF-TEM micrographs of the post-mortem HR850 sample showing (a) slip band formation and intersection, (b) large strain localisation at slip band intersections and (c) interaction between slip bands with $\kappa$-carbides. Red arrows highlight some of the more visible dislocations being pinned by $\kappa$-carbides. Inset: Magnified region within the red square showing dislocations being pinned by rows of $\kappa$-carbides. Beam direction parallel to $\langle001\rangle$.}
	\label{fig:deformation}
\end{figure}

\noindent
where A is a constant equal to 0.4. The dislocation density was therefore found to be $1.2 \times 10^{14}$ m\textsuperscript{-2}, this is in comparison to $10^{10}$ m\textsuperscript{-2} in recrystallised grains and $10^{15}$ m\textsuperscript{-2} in cold worked grains \cite{Cottrell1995}. This shows that recovery at the ageing temperature of 550 \degree C in these alloys is remarkably slow, \textit{i.e.} due to the high Mn content on diffusion rates.

Another significant observation was that the high dislocation density did not alter the $\kappa$-carbide precipitation behaviour in the HR850 sample. In a study by Brasche \textit{et al.} \cite{Brasche2020}, a steel of similar composition was cold rolled and annealed/aged at a temperature between 600-800 \degree C. This led to the precipitation of large $\kappa$-carbides distributed along dislocation networks introduced during the cold rolling process. Similar results were observed in a study by Zhang \textit{et al.} \cite{Zhang2017g}. It was theorised that the enhanced diffusion of solute atoms along dislocation cores facilitated the coarsening of $\kappa$-carbide nuclei in the viscinity of these defects \cite{Brasche2020,Zhang2017g}. However, the aforementioned effect was not observed in HR850 which was still able to form a homogeneous distribution of fine $\kappa$-carbides. It may be theorised that the dislocation substructure of HR850 comprised of an even distribution of Geometrically Necessary Dislocations (GNDs) of low misorientations, \textit{i.e. }$<2\degree$ (Figure \ref{fig:armicro}g) as compared to a cold rolled or cold rolled then recovered steel which contains GNDs of high misorientations, dislocation entanglements and other dislocation structures which were shown to facilitate $\kappa$-carbide coarsening \cite{Brasche2020,Zhang2017g}. 

To investigate the hardening behaviour of HR850, a TEM foil was obtained from the post-mortem tensile sample. From Figure \ref{fig:deformation}, the deformation substructure comprised of intersecting slip bands, forming Taylor lattice structures \cite{Gutierrez-Urrutia2014,Hansen1990} and also a large build up of strain at slip band intersections. The formation of intersecting slip bands with increasing strain was described as Slip band Refinement Induced Plasticity (SRIP) \cite{Welsch2016,Haase2017}. In a fine grained (9 \textmu m) steel, Haase \textit{et al.} \cite{Haase2017} found that slip bands nucleated more rapidly at low strains compared to a coarse grained (110 \textmu m) steel. It was likely that the increased grain boundary area facilitated the rapid initial nucleation and formation of slip bands. This effect was reflected in the higher SHR in HR850 at low strains in Figure \ref{fig:kappa-tensiles}b and shows that the HR850 sample still deforms in a manner typical of this class of steel. 

However, it should be noted that the domain size of HR850 (approximately 480 nm wide) at the failure strain was considerably larger than other steels of similar composition and grain sizes at lower strain levels ($\leqslant 200$ nm wide) \cite{Gutierrez-Urrutia2014,Haase2017,Wu2015a}. It has been proposed that the critical slip band spacing and therefore slip band saturation is reached when long range stresses between slip bands reach a certain limit \cite{Haase2017, Kim2019b}. It was therefore likely that while the high residual dislocation density in HR850 did not appear to impede the formation of slip bands, it would have contributed to the long range stresses between slip bands, resulting in a less refined structure and a larger domain size.

Finally, it is of interest that dislocations were observed piling up behind rows of $\kappa$-carbides in Figure \ref{fig:deformation}c, lending support for Equation \ref{eq:ppt_strength} as proposed by Yao \textit{et al.} \cite{Yao2017}.

In summary, it was shown that by simply lowering the rolling temperature from 1100 \degree C to 850 \degree C, no adverse effects such as ferrite formation, irregular and/or intergranular precipitation of $\kappa$-carbides and change in deformation mechanisms were observed. Instead, the grain size was refined from 72.2 \textmu m to 14.2 \textmu m and the yield strength was improved by 299 MPa. It is hoped that these findings will help develop industrially feasible thermomechanical processing routes that further strengthen $\kappa$-carbide strengthened steels.

\section{Acknowledgements}
TWJK gratefully acknowledges the provision of a studentship from A*STAR, Singapore. KMR and DD would like to acknowledge funding from EPSRC, United Kingdom, grant number EP/L025213/1. VAV would like to acknowledge funding from Rolls-Royce plc and Imperial College London under the Junior Research Fellowship scheme.

\bibliographystyle{ScriptaThomas}
\bibliography{Library}

\end{document}